# Kagome goldene with flat bands and Dirac nodal line fermions via line-graph epitaxy

Qiwei Tian[1, 2], Sahar Izadi Vishkayi[3], Chen Zhang[1], Jiang Zeng[1], Bo Li[1], Li Zhang[1], Long-Jing Yin[1], Yuan Tian[1], Meysam Bagheri Tagani[3,4*], Lijie Zhang[1, 2*], and Zhihui Qin[1*]

1. *Key Laboratory for Micro/Nano Optoelectronic Devices of Ministry of Education & Hunan Provincial Key Laboratory of Low-Dimensional Structural Physics and Devices, School of Physics and Electronics, Hunan University, Changsha 410082, China*
2. *Research Institute of Hunan University in Chongqing, Chongqing 401120, China*
3. *International Research Centre Magtop, Institute of Physics, Polish Academy of Sciences, Aleja Lotników 32/46, PL-02668 Warsaw, Poland*
4. *Department of Physics, University of Guilan, P.O. Box 41335-1914, Rasht, Iran*

*Corresponding authors:

lijiezhang@hnu.edu.cn; m_bagheri@guilan.ac.ir; zhqin@hnu.edu.cn;

**Abstract**

**The kagome lattice has emerged as a promising platform for investigating exotic quantum phases. However, achieving a single-atomic-layer kagome lattice in elemental materials remains a significant challenge. Here, we introduce line-graph epitaxy, a novel approach that enables the atomic-scale synthesis of goldene, a monolayer of elemental gold atoms arranged in a kagome lattice. Through scanning tunneling microscopy/spectroscopy (STM/STS) and density functional theory (DFT) calculations, we demonstrate the formation of kagome goldene, featuring a nearly flat band, with a characteristic spectral feature approximately 1.1 eV below the Fermi level. Notably, the flat-band-derived spectral feature is strongly suppressed at the zigzag edges of goldene nanoflakes, while enhanced edge-associated spectral weight is observed at positive energies. Furthermore, our calculations show that weak interlayer interactions between goldene and the underlying $Au_2Ge$ substrate generate dual Dirac nodal lines through a proximity effect. These findings offer not only a novel strategy for constructing elemental kagome lattices, but also a generalizable framework for fabricating and controlling line-graph materials.**

The kagome lattice, recognized as the line graph of the honeycomb lattice, provides an ideal platform for investigating geometric frustration, topological electronic states, and strong electron correlations [1-7]. Although numerous bulk materials contain kagome layers [8-15], these are typically coupled with non-kagome atoms or layers, complicating the isolation of their intrinsic quantum properties. The distinctive electronic structure of the kagome lattice—characterized by flat bands and Dirac excitations—offers fertile ground for discovering correlated topological phases [16-22]. Recent efforts to engineer such electronic features have relied on twisted superlattices, alloying, or surface patterning [23-25]; however, these strategies often involve multi-component materials or yield only imperfect kagome characteristics. The synthesis of a chemically pure, long-range-ordered kagome monolayer thus remains a major challenge [26,27]. Two-dimensional elemental metals [28,29]—particularly gold, known for its chemical stability and rich surface properties [30-33]—present a promising alternative. Although monolayer gold ("goldene") has been recently realized [34-37], a kagome-structured version has not been achieved until now.

In this Letter, we report the first realization of a single-element kagome monolayer—kagome goldene—constructed via line-graph epitaxy, which exploits the geometric template of the $Au_2Ge$ substrate to deterministically arrange Au atoms into a kagome lattice. In contrast to previously reported kagome platforms based on multi-component systems or artificial patterning, this approach yields a chemically pure, monoelemental kagome network whose electronic states derive directly from lattice geometry. Combining scanning tunneling microscopy/spectroscopy (STM/STS) and density functional theory (DFT) calculations, we identify kagome-derived electronic features including a nearly flat band with a characteristic spectral feature around −1.1 eV. We observe edge-induced electronic modifications in kagome goldene nanoflakes. Furthermore, our calculations reveal dual Dirac nodal lines (DNLs) arising from weak proximity coupling to $Au_2Ge$. Layer-resolved calculations further show that these kagome-derived states remain energetically and spatially distinct from substrate bands, indicating weak interlayer hybridization. Together, these results establish kagome goldene as a chemically pure, symmetry-controlled platform for geometry-driven electronic phenomena.

Figure 1a illustrates the construction of the kagome lattice from a triangular $Au_2Ge$ substrate, where gray (A, Ge) and black (B, Au) atoms form the template, and red dots (C) represent the epitaxial Au atoms constituting the kagome layer. The full stacking sequence—kagome goldene, $Au_2Ge$, and Au(111)—is depicted in Fig. 1b. Determining the epitaxial relationship between this topmost layer and the underlying $Au_2Ge$ is essential [38]. An STM image of the clean $Au_2Ge$ surface is shown in Fig. 1c along with the structural model (inset). After annealing, Au atoms segregate to the surface and self-assemble into a kagome lattice. Figure 1d displays an STM image of a resulting kagome island, with an apparent height difference of ~230 pm relative to the substrate (Fig. 1e), consistent with a single atomic layer. To exclude Ge termination, we performed DFT calculations of a putative Ge kagome layer on $Au_2Ge$. This structure relaxed into a triangular lattice with a Ge–Ge bond length of ~2.78 Å and non-coplanarity (~0.4 Å vertical displacement), inconsistent with a kagome geometry (see Fig. S1 in Supplementary Material [39]). We thus propose a model with an Au kagome layer atop $Au_2Ge$. Figures 1f and 1g show high-resolution STM images confirming the kagome structure, further supported by fast Fourier transform (FFT) analysis (Fig. 1h and Fig. S2) and real-space line profiles (inset of Fig.

1f). The dominant six-fold symmetric peaks at $q_1$ shown in Fig. 1h correspond to the first-order spatial-frequency components associated with the underlying triangular Bravais lattice. The measured peak position is $q_1$=(0.229±0.005) Å$^{-1}$, corresponding to a real-space spacing of approximately 4.37 Å, in good agreement with the expected value $\sqrt{3}a_0$ = 4.30 Å, where $a_0$=2.48 Å is the nearest-neighbor Au–Au distance. Additional Fourier components are observed at $q_2 = \sqrt{3}q_1$ and $q_3$=2$q_1$. Similar Fourier-space features have been reported in previous STM studies of kagome materials [40,41]. We use the Fourier components as complementary structural evidence, while refraining from assigning a specific microscopic origin to the additional components.

The formation of a gold kagome network on $Au_2Ge$ requires the preferential occupation of Au–Au bridge sites by the adatoms. To understand this site selectivity, we compared the relative energies of multiple stacking configurations generated by shifting the kagome layer relative to the substrate along lattice vectors $\boldsymbol{a}_1$ and $\boldsymbol{a}_2$. Configurations with ($\boldsymbol{a}_1$, $\boldsymbol{a}_2$) = (0,0), (1,0), (0,1), and (1,1) correspond to alignments where the kagome layer is centered on the Au–Au bridge sites. As shown in Fig. 2a, the most stable configuration—where the kagome layer is aligned with the bridge sites—is energetically favored by up to 700 meV compared to other stackings. In this optimal configuration (Fig. 2b), the interlayer vertical spacing is 2.8 Å, larger than the apparent STM height of ~2.3 Å (Fig. 1e), reflecting the electronic origin of STM contrast rather than pure geometric spacing. The in-plane registry places Ge atoms at the centers of the kagome hexagons and Au atoms at the triangle centers. The nearest interlayer Au–Au distance is 3.16 Å in this configuration, compared to 2.91 Å in the least stable case (Fig. 2b). Deviation from the high-symmetry registry reduces stability and disrupts the electronic signature of the kagome lattice. As illustrated in Fig. 2d–e, unfavorable stackings induce strong hybridization between the kagome layer and the substrate, suppressing characteristic features such as the flat band, van Hove singularity, and Dirac points. In contrast, the bridge-site configuration preserves a sharp peak in the density of states, marking the presence of the flat band. The dynamic stability of an Au adatom at the bridge site was confirmed using ab initio molecular dynamics (AIMD) simulations at 500 K for 4 ps, during which the atom vibrated locally but did not leave the bridge site (Fig. S3).

To gain deeper insights into the obtained structure, we conducted DFT calculations to examine the electronic properties of both free-standing and substrate-supported kagome goldene. Based on the model described above, we examine a single atomic layer of Au forming a kagome lattice on a monolayer of $Au_2Ge$, where each Ge atom in the bottom layer is positioned at the center of the hexagons. The optimized bond length in the $Au_2Ge$ layer is 2.78 Å, whereas the bond length is 2.40 Å in the kagome lattice, which are in agreement with the experiments as indicated in the line profiles in the inset of Fig. 1f. The $Au_2Ge$ layers are nearly flat, with a corrugation of approximately 5 pm, while the kagome lattice remains completely flat. Figure 3a presents, side by side, the relaxed atomic structure model, the simulated STM image, and the experimental atomic-resolved STM image of the kagome goldene. The simulated STM results are in excellent agreement with the experimental observations, confirming the consistency between the experimental and theoretical models.

We now investigate the electronic properties of the kagome goldene monolayer. When spin-orbit coupling (SOC) is considered, as depicted in Fig. 3b, a nearly flat band is maintained along the M-K path, and type-I and type-II Dirac points are observed. In addition, van Hove singularities are also clear. It is important to note that the nearly flat band and van Hove singularities occur in nearby energy ranges and contribute to distinct features in the calculated density of states (DOS). Dirac cones and nearly flat bands are characteristic features of the kagome lattice [42]. The band structure of the kagome goldene differs from that of the ideal kagome lattice, a lattice with three sites and one orbital for each site, in two aspects. First, the band is not completely flat, but has a finite bandwidth in a part of the first Brillouin zone. Second, there is a gap between the band and the two bands forming the Dirac cone. These features can be attributed to the effects of the d-orbitals of Au atoms, the periodic potential from the $Au_2Ge$ layer, and the second-neighbor hopping in the kagome lattice. The band structure of the sample without SOC is shown in Fig. S4 to highlight how SOC can modulate the electronic properties of the kagome goldene due to heavy Au atoms. The band structure reveals nearly flat bands, with Dirac points present at both positive and negative energy levels. Additionally, van Hove singularities appear at negative energy levels. Comparison of the results reveal that although SOC has a dominant role on the electronic properties of the monolayer, main features of the kagome lattice is preserved.

We then perform STS measurements on the kagome goldene to probe its electronic properties. Figure 3c shows differential conductance ($dI/dV$) spectra measured along the line indicated in the inset, revealing a spatially uniform local density of states (LDOS) across the kagome layer. In addition to the three well-defined occupied-state features discussed below, weak, nearly periodic modulations are observed on the positive-energy side. The origin of these weak modulations is not established by the present data, and they are therefore not considered in the assignment of the kagome-derived electronic states.

Representative spectra acquired at arbitrary positions together with the calculated DOS are shown in Fig. 3d. Three prominent features are observed on the occupied side at approximately −0.3 V, −0.6 V, and −1.1 V, labeled as α, β, and γ, respectively. The pronounced feature at −1.1 V (γ) is associated with the flat-band-derived states, consistent with the calculated band structure and real-space LDOS. The α and β features are further analyzed using band-resolved and momentum-resolved calculations (Figs. S5 and S6). The α feature arises from contributions of multiple nearly degenerate bands around the Γ point, whereas the β feature is dominated by saddle-point states near the M point, consistent with a van Hove singularity. These results indicate that neither feature should be regarded as a direct signature of a Dirac crossing. The $dI/dV$ maps at these characteristic energies exhibit distinct spatial LDOS distributions (Fig. 3e). In particular, the γ feature shows a characteristic row-like pattern, consistent with the calculated flat-band-related LDOS (Fig. S7). The inverse participation ratio (IPR) analysis provides additional quantitative characterization of this sublattice-selective wave-function distribution (Fig. S8), without implying real-space localization of the Bloch state. The Fourier transforms of the $dI/dV$ maps at α, β, and γ are consistent with the kagome lattice periodicity.

Our experimental investigations have confirmed the presence of nanoislands with a kagome structure (Fig. 4a and its inset). A special type of zigzag edge is found as shown in Fig. 4b. STS measurements show that the flat-band-derived spectral feature is strongly suppressed near the zigzag edge, while enhanced edge-associated spectral weight appears at positive energies (Fig. 4c and Fig. S9). The same qualitative spectral redistribution is observed at the boundaries of different kagome goldene samples, as shown in Fig. S10, providing evidence for the reproducibility of this edge-associated electronic modification.

The electronic states detected on these nanoislands diverge markedly from those on the extended kagome sheets. We compared the DOS from each gold atom on nanoislands with that from an ideal kagome lattice atom, as illustrated in Fig. 4d. It is evident that finite-size effects and edge termination modify the electronic properties relative to those of a complete sheet. The calculated nanoisland electronic structure is qualitatively consistent with this experimentally observed spectral redistribution. In these cases, the flat-band-derived spectral feature is strongly suppressed, and additional spectral weight emerges at positive energies near the Fermi level. The calculated wavefunction amplitudes (Fig. 4e) show enhanced weight near the zigzag edges at positive energies, with the intensity decreasing toward the island interior. In contrast, states at negative energies exhibit enhanced weight within the triangular motifs of the kagome lattice. These results highlight the strong sensitivity of the electronic structure to edge geometry in kagome nanoislands.

Examination of the STS results shows that, relative to the kagome-sheet interior, the edge region exhibits enhanced spectral weight around 0.5 eV, while the ~-1 eV feature is suppressed. To further investigate this behavior, we analyze the edge of the semi-infinite kagome structure using surface Green's functions and Wannier functions, employing the WannierTools computational code [43]. Comparing the bulk and ribbon-edge band structures as shown in Fig. 4f, the interior kagome structure exhibits numerous electronic states around -0.8 eV. In contrast, at the ribbon edge, we observe the emergence of electronic states at 0.4 eV, while the states at -0.8 eV are suppressed. Additionally, the Dirac cones associated with the bulk structure are absent in the ribbon-edge electronic structure.

We then consider the heterostructure formed by kagome goldene and $Au_2Ge$. On a larger scale, it is worth noting that there is a much larger periodicity corresponding to the inner dots in the FFT image shown in Fig. 1h, indicating the presence of a moiré pattern, which is visible in the large-scale STM image, as shown in the inset of Fig. 5a. We mark the different stacking sites of the moiré pattern as TOP, hexagonal close-packed (HCP), and face-centered-cubic (FCC), respectively. Since the kagome goldene layer is located on top of the $Au_2Ge$ surface, the periodicity of the moiré pattern is approximately 4.05 nm (inset Fig. 5a). The formation mechanism of the moiré pattern is illustrated in Fig. 5b, where side- and top-view images of the stacking of goldene on $Au_2Ge$, supported by the Au(111) substrate, are clearly shown. The calculated band structure of the entire system is shown in Fig. 5c. As can be seen, there are three bands crossing the Fermi level, with four Dirac points observed along the M-Γ and K-Γ directions around the Fermi level. Additionally, a Dirac cone is observed along the M-Γ and K-Γ direction at an energy of approximately -0.4 eV. Some nearly flat bands also appear around -1.7 eV, below the Dirac cone. The calculated Fermi surface of the heterostructure is shown in Fig. 5d, where two distinct nodal-line–related features are visible along the M–Γ direction, consistent with the corresponding band dispersions. In our previous work, a Dirac nodal line was reported for the isolated $Au_2Ge$ monolayer on Au(111) [38]. Upon stacking kagome goldene on $Au_2Ge$, our calculations show that the nodal line associated with the $Au_2Ge$ layer is preserved, while additional kagome-derived nodal features appear near the K–M region of the Brillouin zone. This coexistence originates from weak interlayer coupling combined with symmetry compatibility and orbital selectivity (in-plane Au $d_{x^2-y^2}$,$d_{xy}$ vs out-of-plane $d_{xz}$,$d_{yz}$) (see Fig. S11). The interlayer interaction is sufficiently weak to prevent strong hybridization or gap opening, allowing nodal features associated with each layer to remain distinguishable. Layer-projected band structures (Fig. 5e-f) demonstrate the energetic separation between kagome-derived and substrate-derived states. SOC-included band calculations (Fig. S12) and layer-resolved Fermi surfaces (Fig. S13) further show that kagome-derived nodal features persist under SOC, whereas the $Au_2Ge$-derived nodal line exhibits stronger SOC sensitivity. These results indicate that the calculated proximity effect reflects orbital-selective band alignment and weak interlayer coupling, which allow distinct nodal-line-related features associated with the two layers to remain identifiable. We note that the identification of nodal-line–related features is based on first-principles calculations and layer-resolved analyses, rather than direct experimental resolution in STS.

In summary, we realize a monolayer of kagome goldene on $Au_2Ge$ via line-graph epitaxy. Weak interlayer coupling preserves the DNL of $Au_2Ge$ and induces an additional DNL in the kagome layer. The heterostructure exhibits hallmark kagome features—including flat-band–like states, van Hove singularities, and Dirac dispersions—confirmed by STM/STS and DFT. Edge termination of kagome nanoislands strongly modifies the electronic states. These results establish kagome goldene as a chemically pure, symmetry-controlled platform for exploring geometry-driven quantum phenomena and highlight line-graph epitaxy as a general strategy for constructing geometrically engineered lattice systems.

**Data availability**

The authors declare that the data supporting the findings of this study are available within the paper and its Supplementary Materials files. The data that support the findings of this study are also available from the corresponding authors upon request.

**Acknowledgements**

This work was supported by the National Key R&D Program of China (2024YFF0727104), the National Natural Science Foundation of China (Grant Nos. 12474167, 12174096, 51972106, 12204164 and 12174095), the Strategic Priority Research Program of Chinese Academy of Sciences (Grant No. XDB30000000), the Natural Science Foundation of Hunan Province, China (Grant Nos. 2025JJ50020, 2025JJ20001 and 2021JJ20026), and the Natural Science Foundation of Chongqing, China (CSTB2025NSCQ-GPX0439).

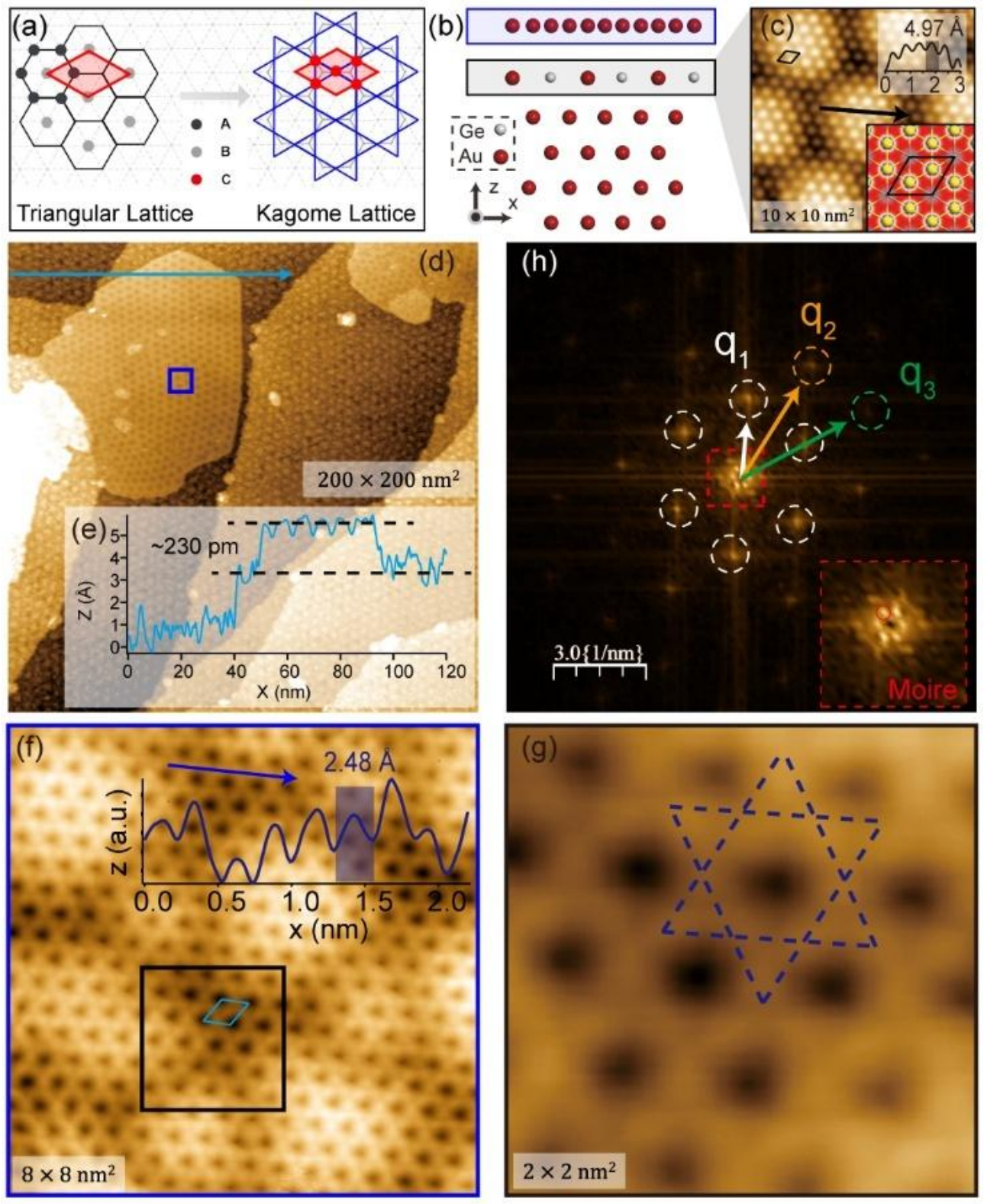

FIG. 1. Synthesis of kagome goldene. (a) Design principle from honeycomb to kagome lattice. (b) Side view of structure model of kagome goldene/$Au_2Ge$/Au(111). (c) Atomic resolution STM image of $Au_2Ge$ with structural model (inset); top inset shows the line profile. (d) Large-scale STM image of kagome goldene on $Au_2Ge$. (e) Apparent height difference (~230 pm) between kagome layer and substrate. (f-g) High-resolution and zoomed-in STM images of the kagome lattice; insets show line profiles. (h) FFT of the STM image in (f). Scanning conditions: (c) $V_S$ = -10 mV, $I_T$=1000 pA; (d) $V_S$ =-100 mV, $I_T$=1000 pA; (f-g) $V_S$ =-100 mV, $I_T$=1000 pA.

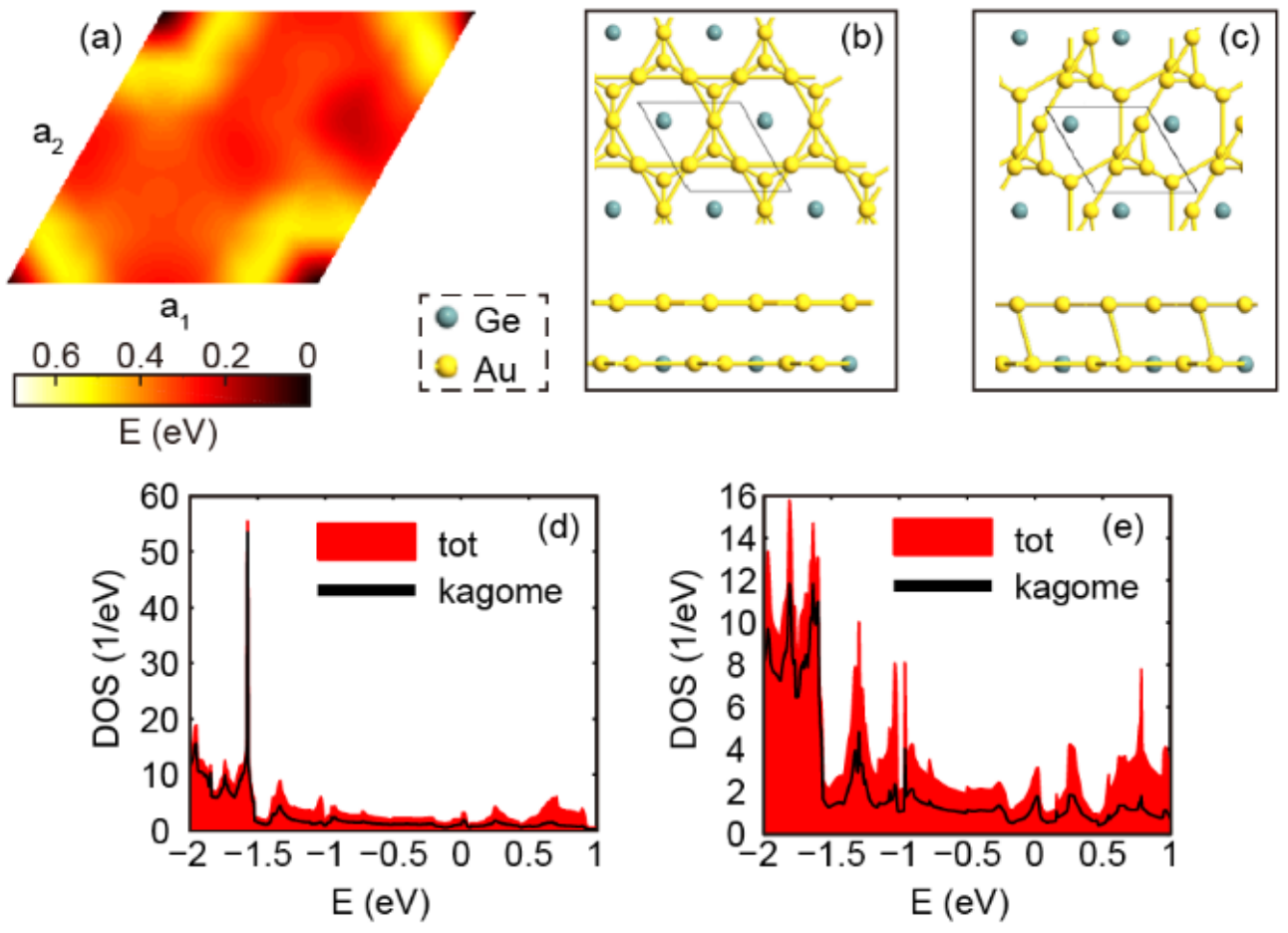

FIG. 2. Stabilization of kagome goldene. (a) Relative energies of different kagome/$Au_2Ge$ stackings obtained by lateral shifts along lattice vectors $\boldsymbol{a_1}$ and $\boldsymbol{a_2}$. (b,c) Top and side views of the most stable and a less stable stacking configurations. (d,e) Total and projected DOS for the stable and unstable configurations, respectively.

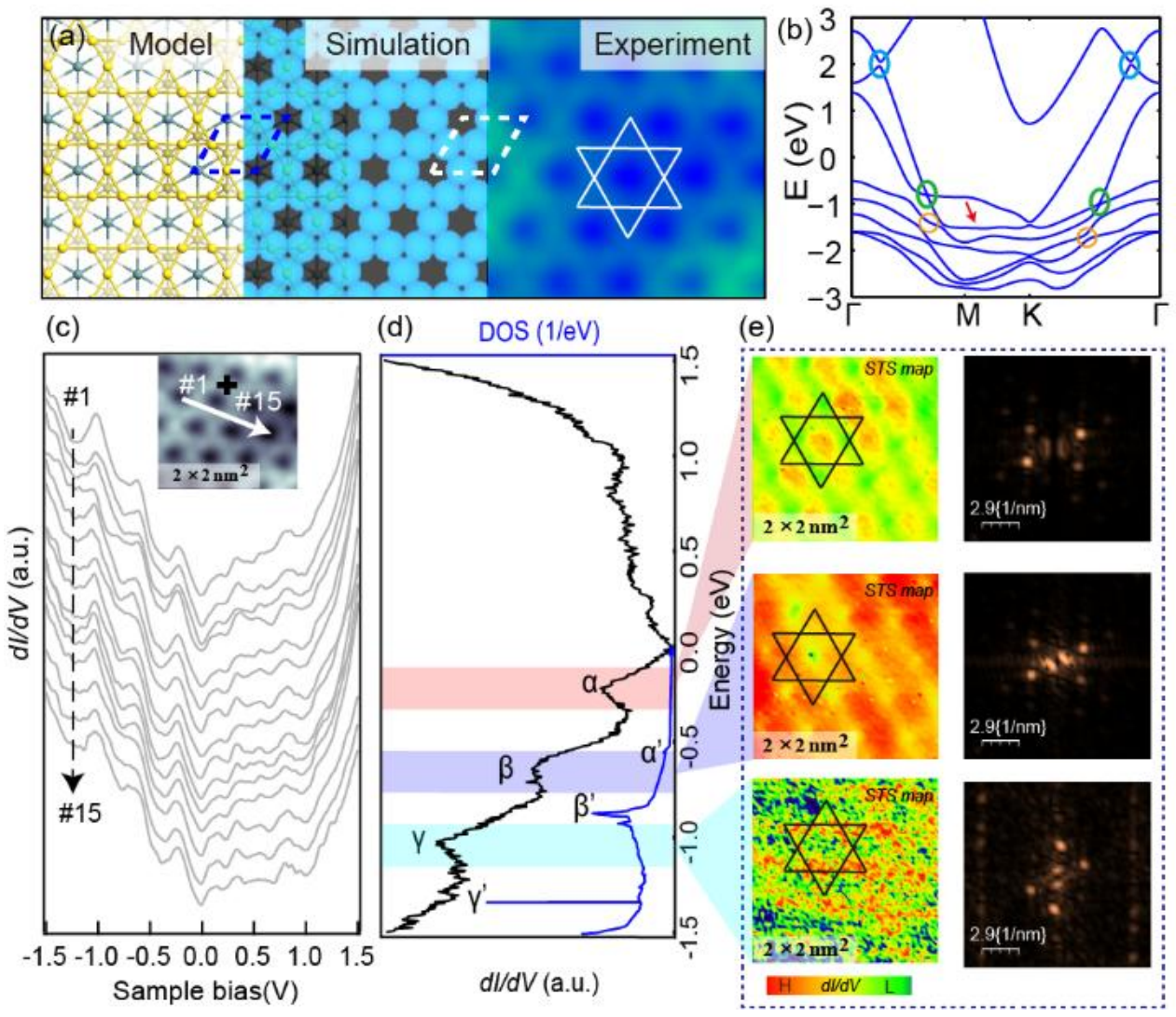

FIG. 3. Electronic structure of kagome goldene. (a) Ball-and-stick model (left), simulated STM image (middle), and experimental STM image (right; $V_s = -100$ mV, $I_t = 1000$ pA). (b) Band structure including SOC. The red arrow points the nearly flat band, and green (blue) circles indicate type-I (type-II) Dirac points. Brown circles show the van Hove singularity points. (c) dI/dV spectra measured along the line marked in the inset STM image. (d) Representative dI/dV spectrum on the kagome lattice compared with calculated DOS. (e) dI/dV maps and corresponding FFTs at −0.3 V, −0.6 V, and −1.1 V.

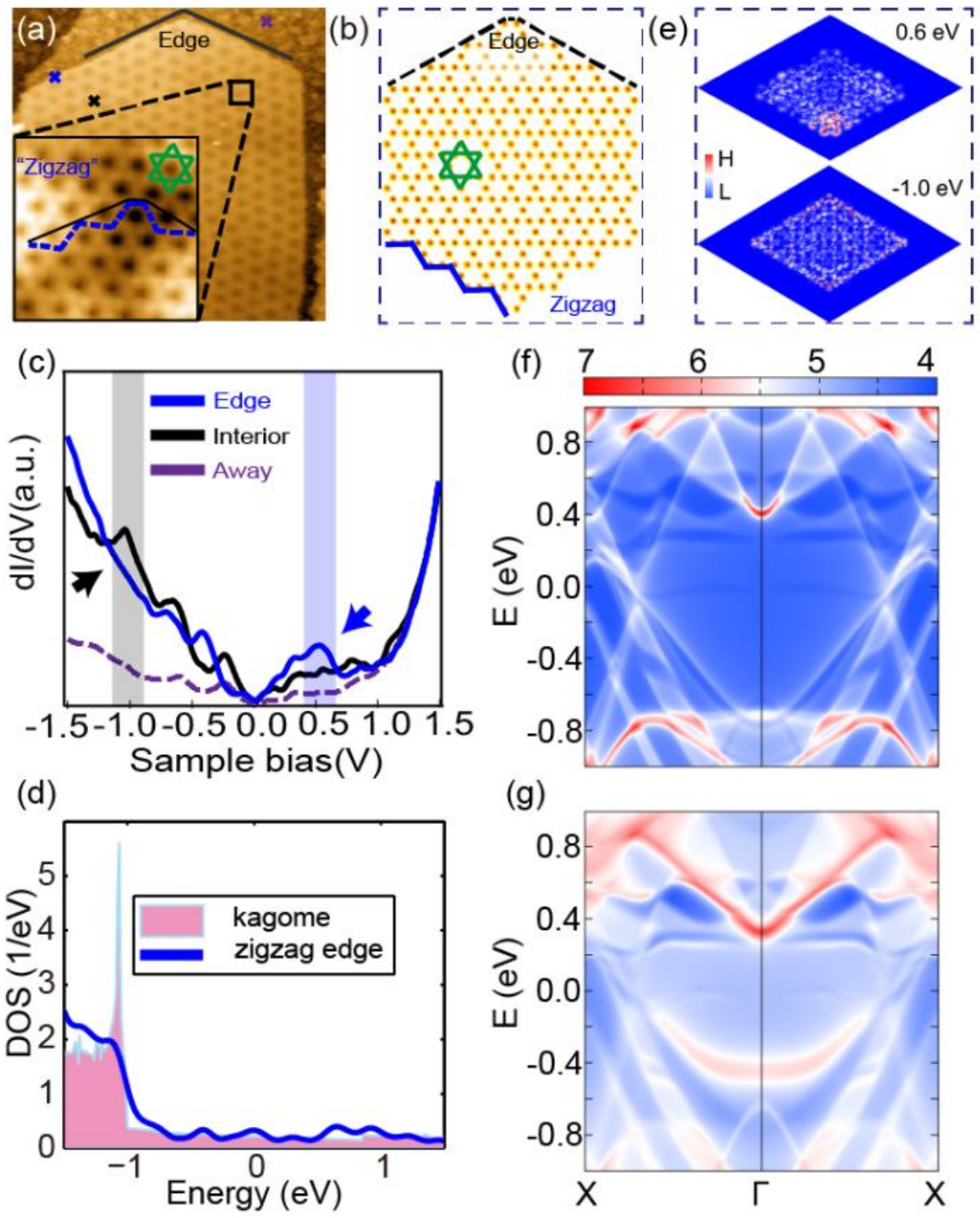

FIG. 4. Edge electronic structure of kagome goldene. (a) STM image of a kagome nanoisland with zigzag edge; inset shows atomic resolution. (b) Top-view atomic structure of a zigzag-edged kagome nanoisland. (c) dI/dV spectra acquired at the edge, interior, and substrate positions marked in (a). (d) Calculated DOS of a perfect kagome sheet and zigzag-edged nanoislands. (e) Calculated real-space wavefunction amplitudes at E = 0.6 eV (top) and −1.0 eV (bottom); color scale indicates relative intensity. (f, g) Band structures of bulk kagome goldene (f) and zigzag-edged ribbon (g).

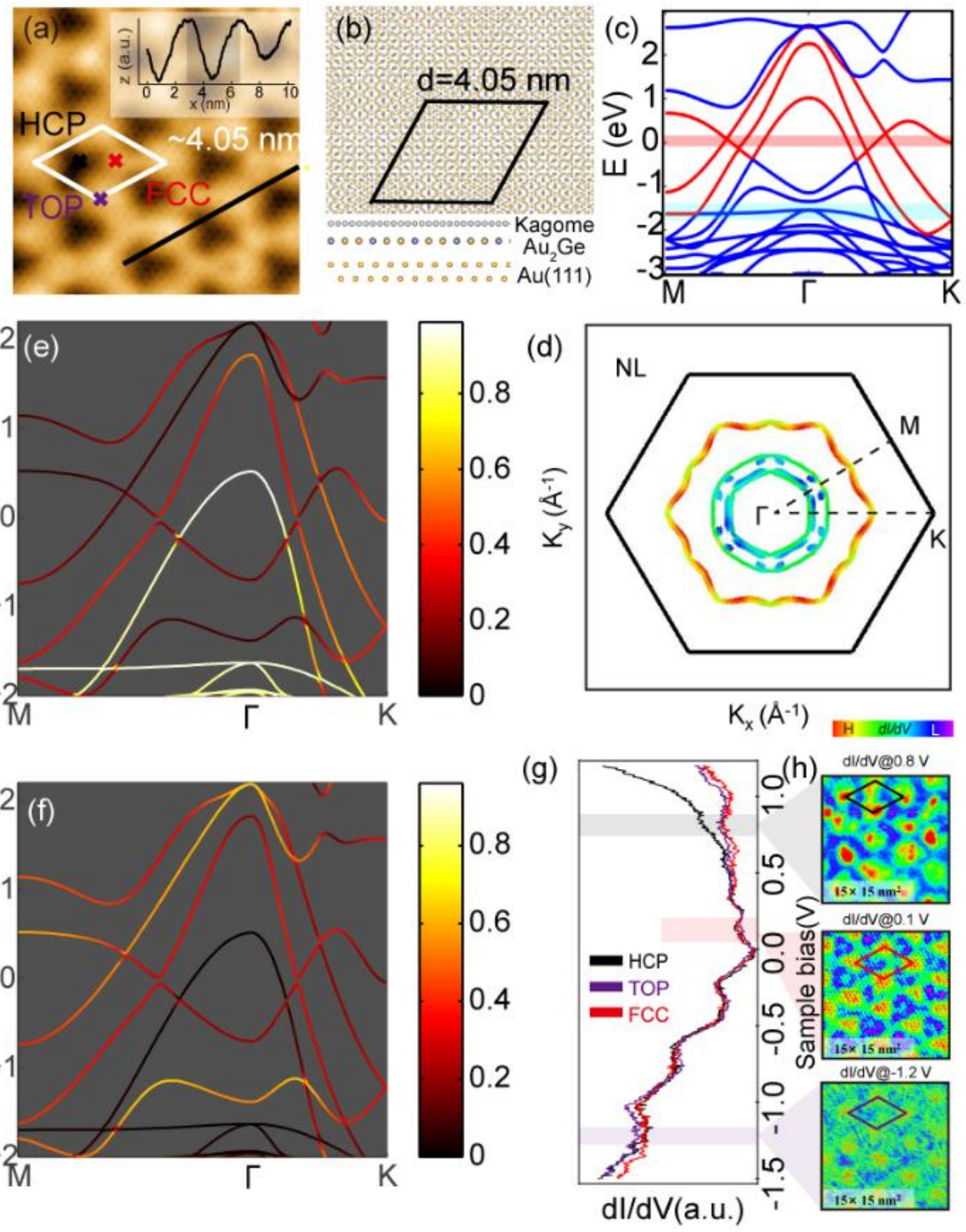

FIG. 5. Kagome goldene/$Au_2Ge$ heterostructure. (a) STM image showing the moiré pattern of kagome goldene on $Au_2Ge$/Au(111) ($V_s = -40$ mV, $I_t = 500$ pA); inset shows the line profile. (b) Top and side views of the structural model. (c) Calculated band structure of the heterostructure. (d) Calculated Fermi surface. (e, f) Layer-projected band structures of kagome goldene (e) and $Au_2Ge$ (f). (g) dI/dV spectra acquired at HCP, TOP, and FCC regions marked in (a). (h) dI/dV maps at −1.2 V, 0.1 V, and 0.8 V.